\documentstyle[twoside,fleqn,espcrc2,psfig]{article}

\newcommand{\AmS}{{\protect\the\textfont2
  A\kern-.1667em\lower.5ex\hbox{M}\kern-.125emS}}
\newcommand{\BS}{{{\it Beppo}SAX}\ \ignorespaces}
\begin{document}
\title{\BS observations of 3C 273}

\author{F.~Haardt\address{Dipartimento di Fisica, Universit\`a 
di Milano, MI} on behalf of the "Flat Blazars" collaboration$^*$}\thanks{
G.~Fossati, P.Grandi, A.~Celotti, 
A.~Treves,               
G.~Ghisellini, L.~Maraschi, G.~Tagliaferri, 
L.~Bassani, F.~Frontera, G.~Malaguti, L.~Nicastro, E.~Pian,
L.~Chiappetti, E.G.~Tanzi, 
S.~Giarrusso, 
M.~Guainazzi, 
P.~Padovani, 
C.~Perola, 
C.~Raiteri, M.~Villata 
M.~Salvati, 
W.~Brinkmann,
W.~Paciesas, 
R.~Sambruna, 
M.~Sikora, 
R.~Staubert, 
M.C.~Urry, 
S.~Wagner. 
}

\begin{abstract}
We present preliminary results of \BS AO1 Core Program observations of 3C 273
performed in January 1997. The source was observed 4 times, from January 13th
to January 23nd, with typical on--source time for the LECS and MECS of $\simeq
12$ and $\simeq 25$ ksec respectively, and of $\simeq 11$ ksec for the PDS. We
also present a close comparison with data obtained during the satellite SVP, in
July 1996. 

On average, the AO1 flux is about a factor $\sim 1.8$ higher than the flux
detected during the SVP, and roughly on the middle of the historical X--ray
flux range. Power law fits with galactic absorption to all observations yield
spectral indices in the range $\Gamma=1.53-1.6$, with the spectrum extending
from 0.2 to at least up to 200 keV without any significant slope change. The
broad band spectrum appears basically featureless, marking a clear difference
from the SVP data, where an absorption feature at low energy and a fluorescence
iron emission line are present. 

The lack of cold/warm matter signatures in our data may indicate that, at this
"high" level of luminosity, the featureless continuum produced in a
relativistic jet overwhelms any thermal and/or reprocessing component, while
the two components were at least comparable during the "low" state of July
1996.
\end{abstract}
\maketitle

\section{3C 273 IN BRIEF}
3C 273 is a nearby ($z=0.158$) quasar, and is one of the extragalactic object
best studied across the entire electromagnetic spectrum. It shows almost all
the features proper of high--luminous quasars, i.e. optical jet with high
polarization, double radio lobes, superluminal motion, variability at all
frequencies, and signs of thermal emission in the UV. The spectral 
energy distribution shows two clear
peaks at UV ($\sim 10$ eV) and $\gamma-$ray ($\sim 1$ MeV) energies. A third
peak is featured at IR energies~[1]. 

Observations in the medium--hard X--ray band (up to 30 keV with {\it GINGA})
show a hard power law continuum with photon index ranging between $1.3-1.6$
~[2]. 
{\it ROSAT} showed evidence of an excess and/or an absorption edge above the
extrapolation of the hard power law at energies $<1$ keV~[3]. 

3C 273 shows prominent $\gamma-$ray emission as well, detected by the 
instruments onboard CGRO. {\it OSSE} data show the
hard power law extending up to $\sim 1$ MeV, with a break at 
higher energy~[4]. 

At energies below 1 keV both an absorption feature
{\it and} a soft excess are present in \BS SVP data~[5]. 

The 3C273 observations discussed here are part of the AO1 Core Program
dedicated to bright blazars. We will also make a comparison of these data with
SVP data. The data reduction presented here has been generally performed with
software released {\it before} September 1997, except that of the LECS
data. Data reduction with the updated software for all the onboard instruments,
and a more detailed analysis, with all the appropriate references, 
will be presented in a forthcoming paper~[6].

\section{ANALYSIS}
The observations of 3C 273 were performed as part of the \BS AO1 Core
Program. The source was observed between Jan. 13th 1997 and Jan. 23rd 1997. In
this period 3C 273 has been observed 4 times, for a total effective exposure of
45.2 ksec in the LECS, 92.1 ksec in the MECS, and 46.3 ksec in the PDS. For
comparison, the exposure times for LECS, MECS and PDS in the SVP were 12, 131
and 64 ksec, respectively. 

\subsection{\bf Temporal}
The \BS data better suited for time analysis are those of the MECS on account
of better statistics and reliable performance stability. 
Combined all the four observations, the MECS count rate monotonically decreases
on time scale of days (Figure~1). Such count rate variation is significant at
99.99\% level. While the first three observations are statistically consistent
with a constant count rate, the last observation is not. The count rate
reverses the decreasing trend, and increases of $\sim 12$\% in about half of a
day. The statistical significance of such variation is 99.88\%. 
Finally, it is important to mention that the average 
AO1 flux was about a factor 1.7 higher than the flux detected during the 
SVP (see Table 1).

\begin{table*}[hbt]
\setlength{\tabcolsep}{1.5pc}
\caption{Single Power Law Spectral Fits: MECS}
\label{tab: MECS}
\begin{tabular}{lccc} 
\hline
\multicolumn{1}{l}{Date(OP)} &
\multicolumn{1}{c}{$\Gamma$} &
\multicolumn{1}{c}{F$_{[2-10]\rm{keV}}$}&            
\multicolumn{1}{c}{$\chi^{2}$/(d.o.f)}\\
&&
\multicolumn{1}{c}{($10^{-10}$ erg/cm$^2$/s)}&\\
\hline
               &                 &                 &        \\ 
18/Jul/96(SVP) & $1.59\pm 0.01$  & $0.69\pm 0.01$  & 102/94 \\ 
               &                 &                 &        \\ 
13/Jan/1997(11)    & $1.56\pm 0.02$  & $1.22\pm 0.01$  &  26/39 \\ 
               &                 &                 &        \\ 
15/Jan/1997(12)    & $1.56\pm 0.02$  & $1.18\pm 0.01$  &  31/39 \\ 
               &                 &                 &        \\ 
17-18/Jan/1997(13) & $1.61\pm 0.02$  & $1.11\pm 0.01$  &  40/39 \\ 
               &                 &                 &        \\ 
22-23/Jan/1997(15) & $1.54\pm 0.02$  & $1.07\pm 0.01$  & 41/39  \\ 
&&&\\
\hline
\multicolumn{4}{@{}p{120mm}}
{Column density fixed at the Galactic value $N_{\rm H}=1.69\times 10^{20}$ cm$^{-2}$.}
\end{tabular}
\end{table*}

\begin{figure}
\psfig{file=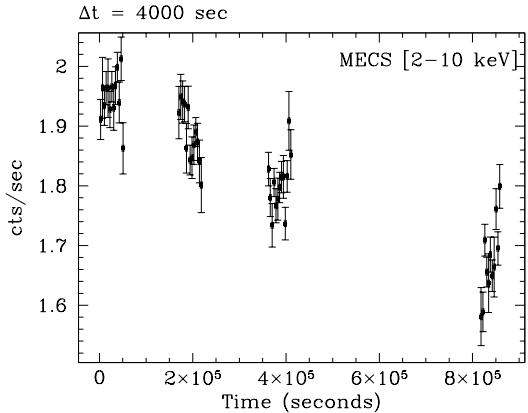,angle=270,rheight=5.0truecm,width=8.1cm}
\caption[h]{MECS light curve. The bin width adopted is 4000 sec.}
\end{figure}

\subsection{Spectral}

MECS data of all the four observations are well described by a single power law
with Galactic absorption ($N_H=1.69\times 10^{20}$ cm$^{-2}$). The energy index
is rather flat, with $\Gamma\simeq 1.55$ for three of the four observations,
consistent with {\it GINGA} results, but slightly flatter than {\it ASCA} 1994
observation~[7]. The spectrum in the third A01 observation (Jan 17) is instead
slightly steeper ($\Gamma=1.61\pm{0.02}$). Similar spectral variability 
has been recently reported, based on {\it ASCA} observations~[7], 
though we can not confirm the anti--correlation of the spectral index 
with the flux seen by {\it ASCA}.

A possible Fe emission line at the expected energy is consistent with zero
flux. There is not indication of any other spectral feature deviating from the
power law in any of the four observations. A summary of power law spectral fits
of MECS data can be found in Table 1. 

Power law fits to the four PDS datasets alone are satisfactory, and
give a spectral index totally consistent with that extrapolated from the fit to
MECS data. Due to the larger error in the determination of the slope, spectral
variations of small amplitude as observed in the MECS data are not detectable.

LECS data were analyzed with the [0.2-4.] keV range using the updated software
released on September 1997, after correction of calibration problems
below 0.7 keV. We checked the presence of absorption features using various
methods, described elsewhere. The main point is that LECS data are consistent
with what derived from higher energy MECS and PDS observations. 

\begin{figure}
\psfig{figure=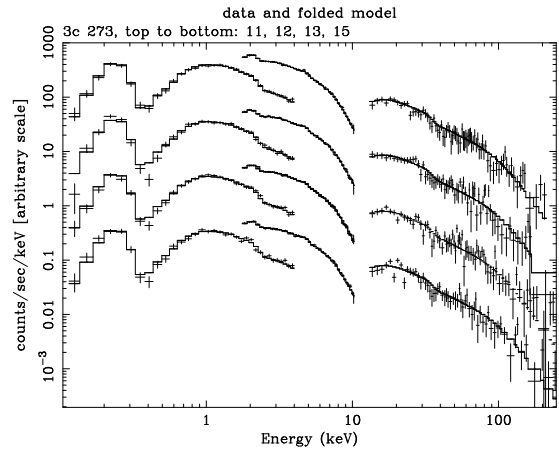,rheight=5.2truecm,width=8.2truecm,angle=270}
\caption[h]{Data for the four observations. The first observation is on top.
For clarity, each successive dataset is scaled in count rate by an order
of magnitude with respect to the previous one.}
\end{figure}

In summary, 3C 273 exhibits a featureless continuum all through the [0.2--200]
keV range, well represented by a single (absorbed) power law (Figure~2). This
in contrast with SVP observations, where the continuum flux level is almost a
factor 2 lower, and an absorption feature is present in the LECS data at $\sim
0.6$ (observer frame) keV, as well a soft excess below 0.3 keV. To check the
reliability of SVP absorption feature and soft excess detection, we computed
the ratio between the SVP and the AO1 LECS counts, as this quantity is
essentially instrumental--effect free. Figure~3 unambiguously shows that the
soft excess and the absorption feature are genuinely present in lower state
observed during the SVP. 

\begin{figure}
\psfig{figure=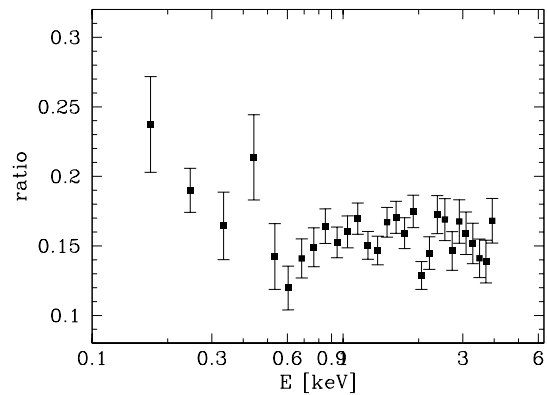,rheight=5.0truecm,width=8.2truecm,angle=270}
\caption[h]{Ratio of SVP/AO1 LECS data.}
\end{figure}

\section{DISCUSSION}
The compared analysis of SVP and AO1 \BS data of 3C 273 led to the following
preliminary scientific results: 

\begin{enumerate}
\item{} The X--ray spectrum is probably the combination of a jet--like
component and a more isotropic disk--like component. The jet--like radiation
passes through partially ionized absorbing material, causing the trough 
observed in SVP data. 
\item{} The light curve and the comparison with SVP data cannot be interpreted
in terms of variable jet--like emission, and a stable disk--like one. In fact,
the disk--like component must have varied since SVP observation, otherwise a
weak soft excess should be present in AO1 data. This means that the jet and
disk emission are both variable, probably not--correlated. 
\item{} The SVP
absorption feature is probably due to highly ionized oxygen along the line of
sight. At the moment, it is not clear if the lack of absorption in LECS AO1
data is consistent with the increased ionization of the absorbing material 
caused by the higher continuum level with respect 
to the SVP. A detailed analysis is in progress. 
\end{enumerate}

This research has made use of SAXDAS linearized and cleaned event files
(rev0.0) produced at the \BS Science Data Center.

\end{document}